\documentclass[conference]{IEEEtran}
\IEEEoverridecommandlockouts
\usepackage{cite}
\usepackage{amsmath,amssymb,amsfonts}
\usepackage{algorithmic}
\usepackage{graphicx}
\usepackage{textcomp}
\usepackage{xcolor}
\def\BibTeX{{\rm B\kern-.05em{\sc i\kern-.025em b}\kern-.08em
    T\kern-.1667em\lower.7ex\hbox{E}\kern-.125emX}}
    \def\bx{{\mathbf x}}

\def\b0{{\mathbf 0}}

\newcommand{\beq}{\begin{equation}}
\newcommand{\eeq}{\end{equation}}

\def\ba{\mbox{\boldmath $a$}}
\def\bb{\mbox{\boldmath $b$}}

\def\bh{\mbox{\boldmath $h$}}
\def\by{\mbox{\boldmath $y$}}

\def\bx{\mbox{\boldmath $x$}}

\def\bs{\mbox{\boldmath $s$}}

\def\bq{\mbox{\boldmath $q$}}

\def\by{\mbox{\boldmath $y$}}

\def\bU{\mbox{\boldmath $U$}}

\def\mB{\mbox{$\mathbf{B}$}}

\def\mH{\mbox{$\mathbf{H}$}}
\def\mI{\mbox{$\mathbf{I}$}}
\def\mL{\mbox{$\mathbf{L}$}}
\def\mY{\mbox{$\mathbf{Y}$}}

\def\mU{\mbox{$\mathbf{U}$}}
\def\mV{\mbox{$\mathbf{V}$}}

\newcommand{\ds}{\displaystyle}

\newcommand{\qedsymbol}{\hspace{\fill}\rule{1.5ex}{1.5ex}}

\usepackage{amssymb}
\begin{document}

\title{Topological Signal Processing over Cell Complexes\\
\thanks{This work was supported in part by H2020 EU/Taiwan Project 5G CONNI, Nr. AMD-861459-3 and in part by MIUR under the PRIN Liquid-Edge contract.}}

\author{\IEEEauthorblockN{Stefania Sardellitti, Sergio Barbarossa, Lucia Testa}
\IEEEauthorblockA{\textit{DIET Dept., Sapienza University of Rome} \\
%\textit{Sapienza University of Rome}\\
\{stefania.sardellitti, sergio.barbarossa, lucia.testa\}@uniroma1.it}
%\and
%\IEEEauthorblockN{Sergio Barbarossa}
%\IEEEauthorblockA{\textit{Dept.  DIET} \\
%\textit{Sapienza University of Rome}\\
%sergio.barbarossa@uniroma1.it}
%\and
%\IEEEauthorblockN{Lucia Testa}
%\IEEEauthorblockA{\textit{Dept.  DIET} \\
%\textit{Sapienza University of Rome}\\
%lucia.testa@uniroma1.it}
}

\maketitle

\begin{abstract}
The Topological Signal Processing (TSP) framework has been recently developed to analyze signals defined over simplicial complexes, i.e. topological spaces represented by finite sets of elements that are closed under inclusion of subsets \cite{barb_2020}.  However,  the same inclusion property represents sometimes a too rigid assumption that prevents the application of simplicial complexes to many cases of interest. The goal of this paper is to extend TSP to the analysis of signals defined over cell complexes, which represent a generalization of simplicial complexes, as they are not restricted to satisfy the inclusion property. In particular, the richer topological structure of cell complexes enables them to reveal cycles of any order, as representative of data features. We propose an efficient method to infer the topology of cell complexes from data  by showing how their use 
enables sparser edge signal representations than simplicial-based methods. Furthermore, we show how to design optimal finite impulse response (FIR) filters operating on solenoidal and irrotational signals in order to minimize the approximation error with respect to the desired spectral masks.
\end{abstract}


\section{Introduction}
In the last years, the ever growing interest in machine learning and  complex networks has motivated the development of models and tools to represent and analyze data on topological spaces. The recent field of Graph Signal Processing (GSP) \cite{ortega2018graph} has emerged as a powerful tool for the analysis of signals defined over the vertices of a graph, which is a simple topological space where edges encode {\it pairwise} relationships among data.  
However, in many applications,  for example in gene regulatory networks, the {\it multiway} relationships among complex molecules, like genes, proteins or metabolites, cannot be grasped using only pairwise relationships \cite{lambiotte2019networks}. To overcome  the limitations of graph-based approaches,  a more general topological signal processing framework (TSP), operating over higher-order structures able to capture multiway relations such as simplicial complexes, has been recently introduced in \cite{barb_2020}, \cite{barb_Mag_2020}, \cite{SCHAUB2021}. Simply put,  a topological space is a set $\mathcal{V}$ of elements along with a set of multiway relations among them. An {\it abstract simplicial complex} is an example of topological space represented by a set $\mathcal{S}$ composed of subsets of various order satisfying the inclusion property, so that if a set $\mathcal{B}$ belongs to $\mathcal{S}$ then any subset of $\mathcal{B}$ also belongs to $\mathcal{S}$  \cite{munkres2018elements}. 
%Methods based on simplicial complex have already been applied in a variety of fields, including statistical ranking \cite{jiang2011}, tumor progression analysis \cite{roman2015simplicial}, and control systems \cite{muhammad2006control}. 
Recently, the authors of \cite{SCHAUB2021} provide a comprehensive tutorial on the broad topic of signal processing over hypergraphs and simplicial complexes, whereas \cite{Yang2021} focuses on the design of finite impulse response filters to process signals defined over simplicial complexes. 
 %The application of simplicial filtering in neural networks operating over simplicial complexes was studied in \cite{ebli2020simplicial}, while in \cite{bodnar2021weisfeiler} the authors presented a message passing neural architecture for simplicial complexes.
However, in many cases of interest, the inclusion constraint associated with simplicial complexes may be quite limiting. For example, in  biological networks, the presence of a reaction among a group of molecules, e.g. genes or proteins, does not imply that a reaction  occurs also among any subgroup of elements. A more general structure, not constrained to respect the inclusion property, but still retaining the algebraic richness of simplicial complexes, is given by {\it cell complexes} \cite{hatcher2005algebraic}, \cite{grady2010}.  Cell complexes are defined as a collection of cells that are sets of points whose subsets does not necessarily belong to the complex.
Cell complexes have already been applied to model complex systems \cite{mulder2018network}, to solve ranking problems \cite{hirani2010least}, to perform neural network-type computation over rich topological spaces \cite{Hajij} or for  
 hierarchical message passing schemes \cite{bodnar2021weisfeilercell}.
The processing of signals over cell complexes has been recently introduced in  \cite{Schaub2021CC}  where it is shown how to design filters to be used in neural networks defined over cell complexes.
 
Our goal in this paper is to extend the TSP framework of  \cite{barb_2020} to signals defined over cell complexes, proposing an algorithm to infer the structure of the complex from data and a method
to find sparse edge signal representations that enable,  using sampling theory, the reconstruction of the whole set of edge signals from the observation of a subset of samples. Furthermore, we propose a method to design optimal FIR filters for the solenoidal and irrotational components. Finally, we corroborate the effectiveness of the proposed methods with numerical tests, by showing how the use of cell complexes yields a substantial performance  gain with respect to simplicial-based methods.
%%%%%%%%%%%%%%

\section{Introduction to Cell Complexes}
%%%%%%%%%%%%%%%%
In this section we introduce the notion of cell complex (CC) and then we present its algebraic representation. Given a discrete set $\mathcal{V}$ of $N$ vertices $v_i$, $i=1,\ldots,N$, a collection $\Delta$ of non-empty finite subsets of $\mathcal{V}$   is called an abstract simplicial complex if it satisfies the inclusion property, meaning that, for every set $X$ in $\Delta$, and every non-empty subset $Y \subseteq X$, the set $Y $also belongs to $\Delta$. An element of the complex is called a $k$-simplex (or simplex of order $k$)  and it simply denotes a set of $k+1$ elements (vertices). If embedded into a real domain, a simplicial complex is composed of vertices ($0$-simplices), edges ($1$-simplices), triangles ($2$-simplices), and so on.\\
A more general structure, not constrained to respect the inclusion property, but still retaining the algebraic richness of simplicial complexes, is given by the cell complexes \cite{hatcher2005algebraic}, \cite{grady2010}.
An {\it abstract cell complex (ACC)}  \cite{klette2000cell}, \cite{steinitz1908beitrage}, denoted as  $\mathcal{C}=\{\mathcal{S},  \prec_b, \text{dim}\}$, is a set $\mathcal{S}$ of elements
along with a binary relation $\prec_b$, called the bounding relation, and with a dimension function, denoted by $\text{dim}(x)$, that assigns to  each $x \in \mathcal{S}$ a non-negative integer, satisfying the following two properties: 
\begin{enumerate}
    \item  if $x \prec_b y$ and $ y \prec_b z$, then $ x \prec_b z$  follows (transitivity);
    %\item  if $x \prec_b y$ and $ y \prec_b x$, then $ x = y$  follows (antisymmetry);
    \item   if $ x \prec_b y$, then $\text{dim}(x)<\text{dim}(y)$ (monotonicity).
\end{enumerate}
An $n$-cell, denoted by $c^{n}$, is a cell $c$ with dimension $n$; $0$-cells are named vertices.
Given two cells $x, y \in \mathcal{S}$, if $x \prec_b y$, we say that $x$ bounds $y$ and $x$ is called a proper side of $y$. The boundary $\partial x^n$ of an   $n$-dimensional cell $x^n$ is defined  as the set of all cells of dimension less than $n$ that bound $x^n$. An ACC is  $k$-dimensional  if the dimensions of all its cells are less than or equal to $k$. 
%The sides of an abstract cell $x$ are not parts of $x$ so that 
%the intersection of two distinct abstract cells is always empty.
The closed cell including its boundary is denoted by 
 $\bar{x}^n=x^n \cup \partial x^n$. 
Two cells $x,y$ are {\it incident} if $x \prec_b y$ or $y \prec_b x$. 
More specifically, we say that $x$ is {\it lower incident} to $y$ if  $x \prec_b y$ and {\it upper incident} to $y$ if $y \prec_b x$.
To associate a topological structure to an ACC, we need to introduce first the concept of open and closed subsets. 
A subset $U$ of $\mathcal{C}=\{\mathcal{S},\prec_b,\text{dim}\}$ is called {\it open} (closed) in $\mathcal{C}$ iff, for every element $x$ of $U$, all elements $y$ of $\mathcal{C}$  upper (lower) bounding $x$ are also in $U$ \cite{klette2000cell}.
According to the axioms of topology, any intersection of a finite number of open (closed) subsets is open (closed).\\
An ACC can be embedded into a Euclidean space. In such a case, $0$-cells represent vertices, $1$-cells are edges  and $2$-cells are polygons. Then, simplicial complexes are particular cases of cell complexes where $2$-cells are triangles. However, cell complexes do not need to respect the inclusion property of simplicial complexes, so that a polygon with more than three sides cannot be a simplex but it can be a cell. \\
Similarly to simplicial complexes, the structure of a $K$-complex is fully described by the set of its incidence (boundary) matrices  $\mB_k$, $k=1,\ldots,K$, where the $k$-th matrix encodes which $k$-cell is incident to which $(k-1)$-cell.
As with graphs, we need to introduce 
the orientation of a cell complex
by generalizing the concept of orientation of a simplex \cite{grady2010}.
Defining the transposition as the permutation of two elements, two orientations are equivalent if each of them can be recovered from the other through an even number of transpositions. 
To define the orientation of a $k$-cell, we may apply a simplicial decomposition \cite{grady2010}, which consists in  subdividing the cell into a set of internal $k$-simplices, so that,
by orienting a single internal simplex, the orientation propagates to the entire cell. An oriented $k$-cell  may then be represented as $c^k=[c^{k-1}_1,\ldots,c^{k-1}_M]$  where two consecutive $(k-1)$-cells, $c^{k-1}_{i}$ and $c^{k-1}_{i+1}$ shares a common $(k-2)$-cell boundary.  
Then, given an orientation of the cells, 
two $k$-order cells are lower adjacent if they share a common face of order $k-1$ or upper adjacent if both are faces  of a cell of order $k+1$. 
Given an oriented cell complex ${\cal C}$,  the set of its incidences matrices $\mB_k$ with $k=1,\ldots,K$ is defined as follows: 
  \beq \label{inc_coeff}
  B_k(i,j)=\left\{\begin{array}{rll}
  0, & \text{if} \; c^{k-1}_i \not\prec_b c^{k}_j \\
  1,& \text{if} \; c^{k-1}_i \prec_b c^{k}_j \;  \text{and} \; c^{k-1}_i \sim c^{k}_j\\
  -1,& \text{if} \; c^{k-1}_i \prec_b c^{k}_j \;  \text{and} \; c^{k-1}_i \nsim c^{k}_j\\
  \end{array}\right. 
  \eeq
where we use the notation $c^{k-1}_i \sim c^{k}_j$ to indicate that the orientations of $c^{k-1}_i$ and $c^{k}_j$ are coherent and $c^{k-1}_i \not\sim c^{k}_j$ to indicate opposite orientations.

Let us consider a cell complex of order two  $\mathcal{C}=\{\mathcal{V},\mathcal{E},\mathcal{P}\}$
where $\mathcal{V}$, $\mathcal{E}$, $\mathcal{P}$ denote the set of  $0$, $1$ and $2$-cells, i.e. vertices, edges and polygons, respectively. We denote their cardinality by $|\mathcal{V}|=V$, $|\mathcal{E}|=E$ and $|\mathcal{P}|=P$. 
Then, the two incidence matrices describing the connectivity of the complex are $\mB_1 \in \mathbb{R}^{V\times E}$ and $\mB_2 \in \mathbb{R}^{E\times P}$, where $\mB_2$ can be written as
\beq \label{eq:B2_cell}
\mB_2=[\mB_{T},\mB_{Q},\ldots,\mB_{P_{S}} ]
\eeq
where  $\mB_{T}$, $\mB_{Q}$ and  $\mB_{P_{S}}$ indicate the incidences between edges and, respectively, triangles, quadrilaterals, up to polygons with $P_{S}$ sides, where each polygon does not include any internal chord between any pair of its vertices.
An interesting property of the incidence matrices is that $\mB_k \mB_{k+1}=\mathbf{0}, \; \forall k$. 
To describe the structure of a $K$-cell complex, we can use the {\it higher order combinatorial Laplacian} matrices given by  \cite{goldberg2002comb}:
\beq
\begin{split}
& \mL_0=\mB_1\mB_1^T,\\
&\mL_k=\mB_k^T\mB_k+\mB_{k+1}\mB_{k+1}^T \; \; \mbox{for} \; k=1,\ldots,K-1\\
&\mL_K=\mB_K^T\mB_K
\end{split}
\eeq
where $\mL_{k}^{l}:=\mB_k^T \mB_k$ and $\mL_{k}^{u}:=\mB_{k+1} \mB_{k+1}^T$ are  the lower and upper Laplacians, expressing, respectively, the lower and upper adjacencies of the $k$-order cells.
Considering w.l.o.g. the first order Laplacian, i.e.
\beq
\mL_1=\mL_{1}^{l}+\mL_{1}^{u}=\mB_1^T \mB_1+\mB_2 \mB_2^T,
\eeq
it holds: i) the eigenvectors associated  with  the  nonzero  eigenvalues  of  $\mL_{1}^{l}=\mathbf{B}^T_1\mathbf{B}_1$  are  orthogonal  to  the eigenvectors associated with the nonzero eigenvalues of  $\mL_{1}^{u}=\mathbf{B}_{2}\mathbf{B}_{2}^T$ and viceversa; ii)
the eigenvectors  associated with the nonzero eigenvalues $\lambda^1$ of  $\mathbf{L}_1$ are either the eigenvectors of $\mL_{1}^{l}$ or those of $\mL_{1}^{u}$; and, finally, iii)
the  nonzero eigenvalues of $\mathbf{L}_1$ are either the eigenvalues of $\mL_{1}^{l}$ or those of $\mL_{1}^{u}$.
This spectral structure of $\mL_1$ induces and interesting decomposition of the whole space $\mathbb{R}^{E}$, the so-called {\it Hodge decomposition} \cite{Lim}, given by
\beq \label{eq:Hodge_dec}
\mathbb{R}^{E} \triangleq \text{img}(\mB_1^T) \oplus \text{ker}(\mL_1)\oplus \text{img}(\mB_2)
\eeq
where the vectors in $\text{ker}(\mL_1)$ are also in 
$\text{ker}(\mB_1)$ and $\text{ker}(\mB_2^T)$.

\section{Analysis of signals defined over CCs}
In this section we extend the fundamental tools to analyze signals defined over simplicial complexes provided in \cite{barb_2020}  to cell complexes. We focus w.l.o.g. on cell complexes of order $2$. Given a cell complex $\mathcal{C}=\{\mathcal{V},\mathcal{E},\mathcal{P}\}$, the signals on vertices, edges and polygons are defined by the following maps: $\bs^0: {\cal V} \rightarrow \mathbb{R}^V$, $\bs^1: {\cal E} \rightarrow \mathbb{R}^E$, and $\bs^2: {\cal P} \rightarrow \mathbb{R}^P$.
A useful orthogonal basis to represent signals of various order, capturing the connectivity properties of the complex, is given by the eigenvectors of the corresponding higher order Laplacian. Then, generalizing graph spectral theory, we can introduce a notion of Cell complex Fourier Transform (CFT) for signals defined over cell complexes. Let us consider the eigendecomposition
$\mL_k= \mU_k \boldsymbol{\Lambda}_k \mU_k^T$
where $\mU_k$ is the eigenvectors matrix and $\boldsymbol{\Lambda}_k$ is a diagonal matrix with entries the eigenvalues $\lambda_i^k$ of $\mL_k$, with $i=1,\ldots, E$. Then, we define the $k$-order CFT as the projection of a $k$-order signal onto the eigenvectors of $\mL_k$, i.e.
\beq \label{eq:GFT}
\hat{\bs}^k \triangleq \mU_k^T \bs^k. \eeq
A signal $\bs^k$ can then be represented in terms of its CFT coefficients as $\bs^k =\mU_k \hat{\bs}^k.$
Exploiting the Hodge decomposition in (\ref{eq:Hodge_dec}), we may always express a signal $\bs^1$ as 
the sum of three orthogonal components \cite{Lim}, i.e.
\beq \label{eq:s1_dec}
\bs^1=\mB_1^T \bs^0+ \mB_2 \bs^2+\bs^1_H.
\eeq
In analogy to vector calculus terminology, the first component $\bs^1_{irr}:=\mB_1^T \bs^0$ is called the {\it irrotational} component since, using the equality $\mB_1 \mB_2=\mathbf{0}$, it has zero curl, i.e. $\mB_2^T \bs^1_{irr}=\mathbf{0}$, while the second term $\bs^1_{sol}:= \mB_2 \bs^2$  is the {\it solenoidal} component, since its divergence defined as $\mB_1 \bs^1_{sol}$ is zero. Finally, the component $\bs^1_H$ is the {\it harmonic} component since it belongs to $\text{ker}(\mL_1)$ and it has zero curl and zero divergence.

\section{Inference of  cell complexes topology}
Hinging on the algorithm proposed in \cite{barb_2020}, in this section we present a method to infer the topology of a cell complex from data. Our goal is to show that cell complexes enable more sparse signal representations, for a  given signal reconstruction error, than simplicial complexes.  We start with inference of the topology of cell complexes of order $2$ from the observations of a set of $M$ edge signals 
$\mY^1=[\by^1(1), \ldots, \by^1(M)]$.
We assume that the graph topology, i.e. the 0-order Laplacian matrix $\mL_0=\mB_1 \mB_1^T$, is known. Then, since $\mL_1=\mB_1^T \mB_1+ \mB_2 \mB_2^T$, we only need to infer the incidence matrix $\mB_2$.
More specifically, we infer the presence of $2$-cells, i.e. polygons of any order, by selecting the columns $\bb_n$ of the boundary matrix $\mB_2$ for which the total circulation of the observed edge signals along the corresponding $2$-cell is minimum. 
As a first step, we need to check if the upper Laplacian $\mL_{1}^{u}$ is really needed to represent the observed data. Then, since the solenoidal and harmonic signals are the only components depending on $\mB_2$, we first project the observed signals onto the space orthogonal to the irrotational component. Thereby, defining the matrix $\mU_{irr}$ containing as columns the eigenvectors of $\mL_1^{l}$ associated with the non-zero eigenvalues, we compute
\beq
\by^{1}_{sH}(i)=(\mI-\mU_{irr}\mU_{irr}^T) \by^{1}(i), \; i=1,\ldots,M.
\eeq
Denoting with $\mY^1_{sH}=[\by^1_{sH}(1), \ldots, \by^1_{sH}(M)]$ the resulting matrix, 
we measure the energy of $\mY^1_{sH}$ by taking its norm: If the norm falls below a threshold, we set $\mB_2=\mathbf{0}$, otherwise  we proceed with the  inference 
of $\mB_2$.
To do that, we extend the algorithm proposed in \cite{barb_2020},  minimizing the total variation of the observed data along {\it all polygons}.
Indicating with $q_n$ a binary coefficient equal to $1$ (or $0$) if the corresponding cell is present (or not), we have
\beq
\mB_2 \mB_2^T=\sum_{n=1}^{N_c} q_n \bb_n \bb_n^T,
\eeq
where  $N_c$ denotes the number of possible $2$-cells in the complex.
Our goal is to find the entries of the vector $\bq=[q_1,\ldots,q_{N_c}]$ as the solution of the following problem:
\beq \label{eq:prob_inf}
\begin{array}{lll}
\underset{\bq \in \{0,1\}^{N_c}}{\min} & \ds \sum_{n=1}^{N_c} q_n \mbox{tr}(\mY^{1\, T}_{sH} \bb_n \bb_n^T \mY^1_{sH} )\\
\quad \text{s.t.} & \parallel \bq \parallel_0=q^{\star}
\end{array}
\eeq
where the number of components $q^{\star}$  is found in a validation phase. Interestingly, this problem, albeit non-convex, admits a closed form solution.
Defining the nonnegative coefficients
$d_n=\sum_{i=1}^{M} \by_{sH}^{1 \; T}(i) \bb_n \bb_n^T\by_{sH}^{1}(i)$, the optimal solution can be derived by sorting in increasing order the coefficients $d_n$ and selecting the columns of $\mB_2$ corresponding to the indices of the  $q^{\star}$ lower coefficients $d_n$.

\section{Sparse signal representation}
Once we have $\mL_1$, we can find a basis  of the observed edge flows using the eigenvectors of  $\mL_1$. Our goal is to find the optimal trade-off between the sparsity of the signal representation and the data fitting error by solving the following basis pursuit problem:% \cite{Donoho98}:
\beq \label{eq:bas_pur}
\begin{array}{lll}
 \underset{{\bs}^1 \in \mathbb{R}^E}{\text{min}} & \parallel
{\bs}^1\parallel_1   \\
 \; \; \text{s.t.} & \parallel
 {\by}^1 -\mV {\bs}^1\parallel_F \leq \epsilon
 \end{array}
\eeq
 where $\mV=\mU_1$ is the eigenvectors matrix associated with $\mL_1$.
Exploiting the bandlimited property enforced by the sparse representations, 
we   reconstruct the overall set of edge signals from a subset of edge samples using a generalization of the MaxDet greedy sampling strategy in \cite{tsitsvero2016signals} and the recovering rule (49)  in \cite{barb_2020}.
\begin{figure}[ht]
\centerline{\includegraphics[width=6.5cm,height=4.0cm]{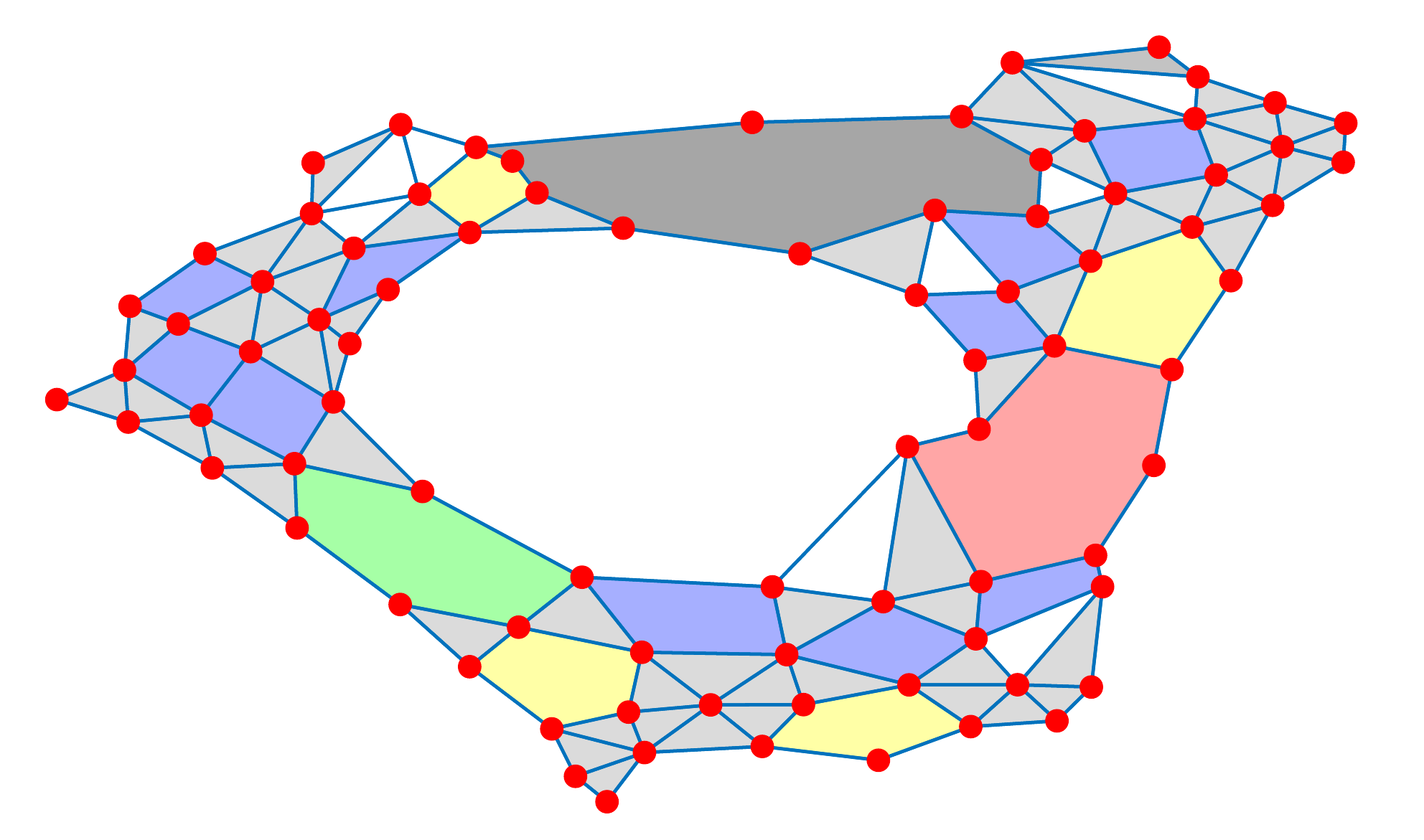}\label{fig_inf_cell}}  \vspace{-0.25cm}
\begin{center}
    \small{(a)}
\end{center}
\centerline{\includegraphics[width=7.5cm,height=4.3cm]{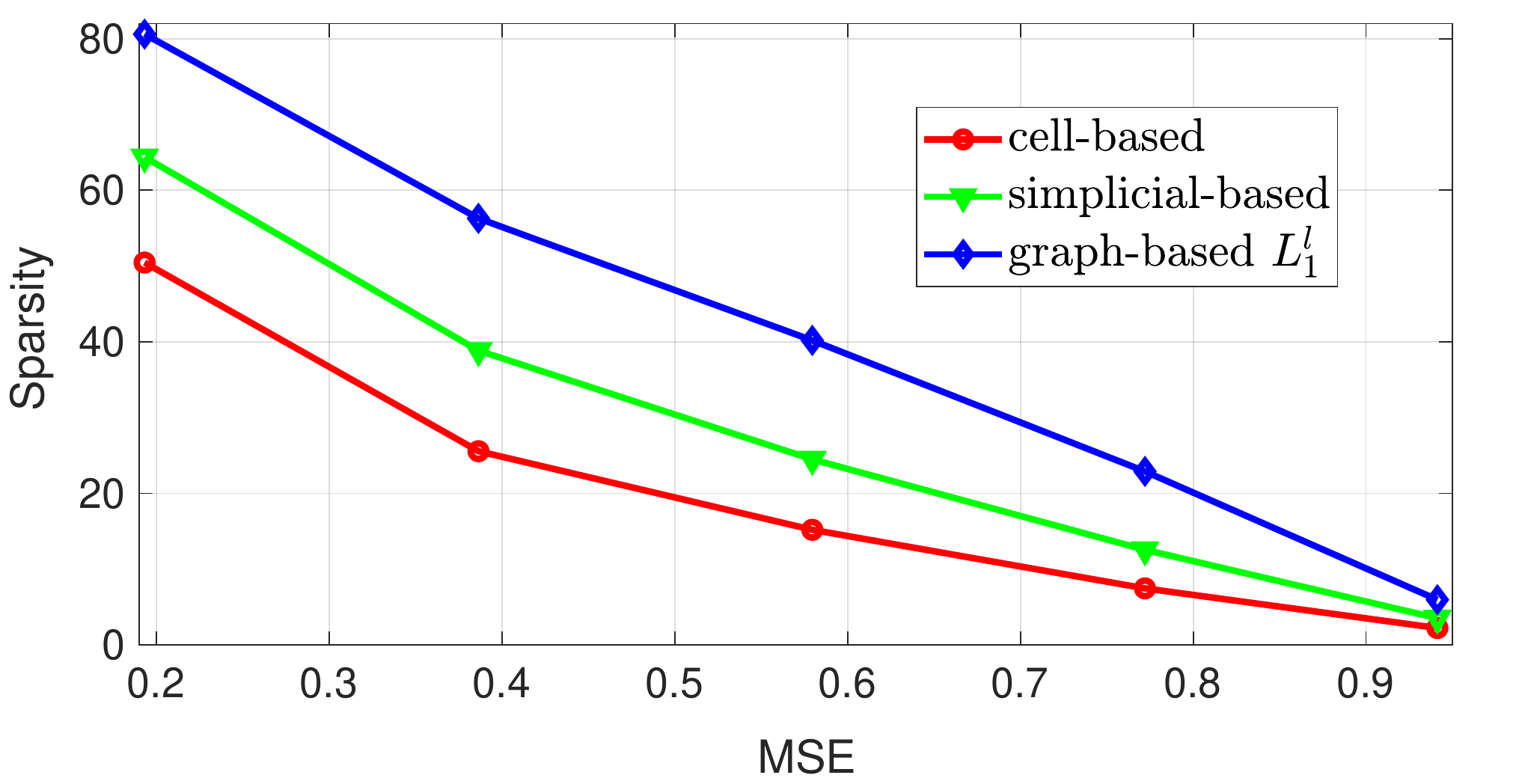}\label{spars_MSE}}%fig_spars_MSE_asilfin.pdf \vspace{-0.2cm}
\begin{center}
   \small{(b)}
\end{center}\vspace{-0.3cm}
\caption{(a) Inferred cell complex; (b) Sparsity vs. mean squared error.} 
\label{fig: fig_MSE}
\end{figure}
To test the effectiveness of the proposed methods, in Fig. $1(a)$ we represent an example of inferred $2$-cell complex from the  observation of $M=100$ independent realizations of edges signals over a complex with $N=82$ nodes, $E=165$ edges and $75$ polygons. Then, we also infer, from the same data-set, a simplicial complex. To compare the different topological spaces, in Fig. $1(b)$ we report the signal sparsity (number of nonzero entries of ${\bs}^1$) versus the mean squared error (MSE) obtained by solving problem in (\ref{eq:bas_pur}), using for $\mV$ the eigenvector matrix associated to $\mL_1^{l}$ and to the first order Laplacians of the inferred simplicial and cell complexes. We can notice from Fig. $1(b)$ that, using a cell-based approach,
we can achieve a better sparsity/MSE trade-off  compared to simplicial based methods, where only triangles are taken into account, or to graph-based methods, where the signal basis is given by the eigenvectors of $\mL_1^{l}$. Then, exploiting the bandlimited property of the signals enforced by their sparse representations, we can use the sampling theory to reconstruct the overall edge signal vector from a subset of edge values, generalizing the approach of \cite{tsitsvero2016signals} to cell complexes. In Fig. $2$ we  illustrate an example of reconstruction, in the presence of noise with variance $\sigma_n^2=0.01$, plotting the  MSE in the signal reconstruction versus the number of  samples used to retrieve the overall signal. From Fig. $2$, we can assess the substantial performance improvement of the cell-based method with respect to graph- and simplicial-based approaches. %\vspace{-0.1cm}
\begin{figure}[t]
\centerline{\includegraphics[width=7.5cm,height=4.6cm]{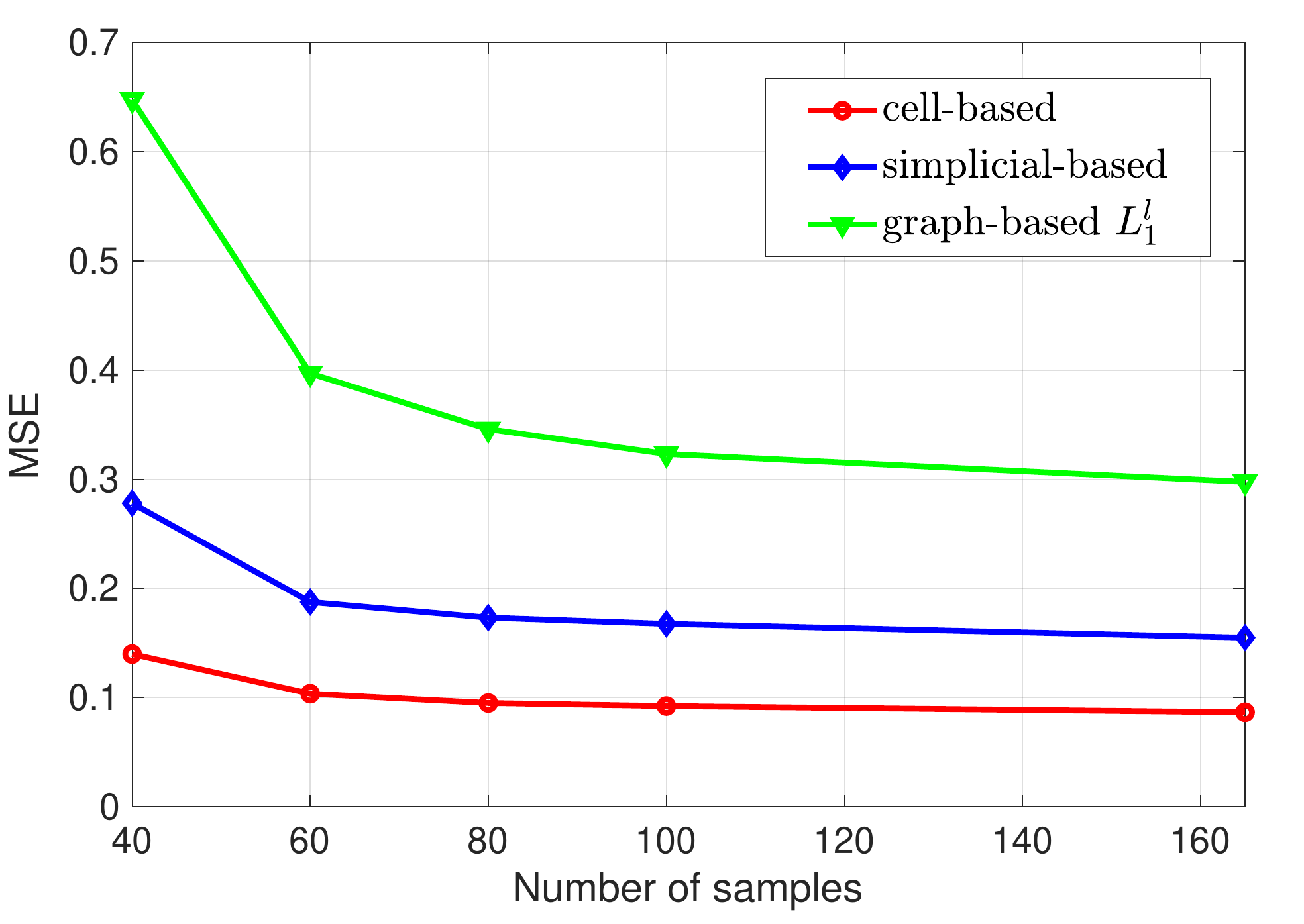}}
\caption{Mean squared error versus number of samples.}
\label{fig: fig_spars}
\end{figure}

%Then, exploiting the bandlimited property enforced by the sparse signal representation, we recover the overall edge signals from the observation of a small set of samples.

\section{FIR filters over cell complexes} 
In this section we propose a method to design FIR filters for the solenoidal and irrotational signals observed over the edges of  cell complexes. 
Graph filters  have been largely investigated in the field of GSP \cite{Moura_14}, \cite{Segarra_17} and their extension to signals defined over the edges of simplicial complex have been considered in \cite{Yang2021}. Recently, filters  defined over cell complexes have been proposed in \cite{Schaub2021CC}. We focus on the design of  FIR filters for the solenoidal and irrotational signals, since the filtering of the harmonic component needs an independent  design as we deeply investigate in  \cite{Sard_Barb_2021}.\\
As a straight generalization of graph filters, a FIR filter  based on the first order Laplacian is a local operator assuming the following form
\beq \label{eq:H0_L1}
\mH= \sum_{k=0}^{K} a_k \mL^{k}_{1} 
\eeq
where $\{a_k\}_{k=0}^{K}$ are the filter coefficients, $K$ is the filter length and  $\mL^{0}_{1}=\mI$. Note that 
the matrices $\mL^{k}_{1}$, $k=1, \ldots, K$ contain information about the neighbors of order $K$, then they represent local operators which linearly combine edge signals from upper and lower neighboring cells. Because of the condition $\mB_1 \mB_2=\mathbf{0}$,  it holds $\mL_1^k=(\mL_{1}^{l})^k+(\mL_{1}^{u})^k$. Thus  
the filtering operation in (\ref{eq:H0_L1}) reduces to the following two orthogonal filters
\begin{equation}
    \label{eq:H0n}
\bx^1=\sum_{k=0}^{K} a_k^{I} (\mL_{1}^{l})^k \bs^1+\sum_{k=0}^{K} a_k^{s} (\mL_{1}^{u})^k \bs^1
\end{equation}
which suggests an independent design of the filters coefficients $a_k^{I}$, $a_k^{s}$ for, respectively, the irrotational and solenoidal components, as also proposed in \cite{Yang2021}. 
Using the CFT defined in (\ref{eq:GFT}), the spectrum of the filter output becomes
\beq \label{eq:FIR_sch}
\begin{split}
\hat{\bx}^1&=\mU_1^T \bx^1=\left( \sum_{k=0}^{K} a_k^{I} \boldsymbol{\Lambda}_{l}^k + \sum_{k=0}^{K} a_k^{s} \boldsymbol{\Lambda}_{u}^k\right) \hat{\bs}^1
\end{split}
\eeq
where we used (\ref{eq:H0n}) by splitting the diagonal matrix $\boldsymbol{\Lambda}_{1}$ containing the eigenvalues of $\mL_1$ into two diagonal matrices $\boldsymbol{\Lambda}_{l}$ and $\boldsymbol{\Lambda}_{u} \in \mathbb{R}^{E \times E}$ containing the eigenvalues of the lower and upper Laplacians, respectively.
%Then, the proposed filters in the spectral domain impose  two different  spectral masks on the irrotational and solenoidal frequencies.
%Note that,  since spectral-based filters are unable to distinguish multiple eigenvalues, we have to assume that the non-zero eigenvalues of $\mL_1$ are distinct. 
It is worth to notice that, by definition of spectral filtering, if the Laplacian $\mL_1$ contains eigenvalues of multiplicity greater than $1$, the filter is not able to distinguish the components belonging to the subspace associated to the multiple eigenvalue. A typical example  occurs when the embedding of the abstract cell complex onto a real domain yields a structure with a number of holes $N_h>1$. In such a case, $\mL_1$ has a null eigenvalue of dimension $N_h$ \cite{munkres2018elements}.
If we rewrite (\ref{eq:FIR_sch}) by distinguishing the term with $k=0$, as
\beq \label{eq:FIRh0}
\hat{\bx}^1=\left( a_0 \mI+\sum_{k=1}^{K} a_k^{I} \boldsymbol{\Lambda}_{l}^k + \sum_{k=1}^{K} a_k^{s} \boldsymbol{\Lambda}_{u}^k\right) \hat{\bs}^1
\eeq
it is evident that all vectors belonging to the kernel of $\mL_1$ are simply scaled by a common coefficient $a_0$.
%The  filtering strategy in (\ref{eq:FIRh0})  has been proposed in \cite{Yang2021} for simplicial complexes.
Note that a coefficient $a_0 \neq 0$ in   (\ref{eq:FIRh0})
has  an adverse effect on the disjoint filtering of the  irrotational and solenoidal signals  since it  lets all the signal components pass through the filter. 
%Furthermore, its effect on the harmonic component is a simple alteration of a scale factor $a_0$ and this calls for a different and independent design of the harmonic filter. 
In this paper, for lack of space, we focus on the filtering of the solenoidal and irrotational signals by designing two independent FIR  filters in which the constant term is removed, so that  they assume the form
\begin{equation}
    \label{eq:H0}
\bx^1=\left(\sum_{k=1}^{K_l} a_k^{I} (\mL^{l}_{1})^k +\sum_{k=1}^{K_u} a_k^{s} (\mL^{u}_{1})^k\right)\bs^1
\end{equation}
where each filter of length, respectively, $K_l$ and $K_u$ imposes a spectral mask on the desired component while filtering out all the others. 
Let us assume that the operators we wish to implement through FIR filters
are given by 
\beq \label{eq:H_filter} 
\begin{split}
\mathbf{H}_{irr}= \mU_{irr} h^I(\boldsymbol{\Lambda}_{irr}) \mU_{irr}, \qquad  \mathbf{H}_{sol}=\mU_{sol} h^s(\boldsymbol{\Lambda}_{sol})\mU_{sol}^T,
\end{split}
\eeq
where $\mU_{irr}$ and $\mU_{sol}$ are the matrices containing  the eigenvectors  associated with the non-zero eigenvalues of $\mL_{1}^{l}$ and $\mL_{1}^{u}$, respectively, and
$\boldsymbol{\Lambda}_{irr}$, $\boldsymbol{\Lambda}_{sol}$ are the  diagonal matrices with entries the corresponding non-zero eigenvalues. The diagonal entries $h^I(\lambda_i)$ and $h^s(\lambda_i)$  of  the matrices $h^I(\boldsymbol{\Lambda}_{irr})$ and $h^s(\boldsymbol{\Lambda}_{sol})$ are the frequency responses of the filters associated with the irrotational and solenoidal eigenvalues.
Our goal is to approximate the operators in  (\ref{eq:H_filter}) with the following   FIR filters 
\beq \label{eq:filter_s_I}
\hat{\mH}_{irr}=\ds \sum_{k=1}^{K_l} a_k^{I} (\mL_{1}^{l})^k, \quad \hat{\mH}_{sol} = \ds \sum_{k=1}^{K_u} a_k^{s} (\mL_{1}^{u})^k.
\eeq
The  filters coefficients $\ba^{I}=(a_k^{I})_{k=1}^{K_l}$ and $\ba^{s}=(a_k^{s})_{k=1}^{K_u}$ can be found solving the following least-squares problems  in the spectral domain 
\beq \label{eq:H_irr1}
\underset{\mathbf{a}^{I} \in \mathbb{R}^{K}}{\min} \quad \parallel \mathbf{h}^{I}-\boldsymbol{\Phi}_{I}  \ba^{I}\parallel^2_F \quad \text{and} \quad \underset{\mathbf{a}^{s} \in \mathbb{R}^{K}}{\min} \quad \parallel \mathbf{h}^s-\boldsymbol{\Phi}_{s}  \ba^{s} \parallel^2_F
\eeq
where  $\mathbf{h}^{I}:=\text{diag}(h^I(\boldsymbol{\Lambda}_{irr})), \mathbf{h}^{s}:=\text{diag}(h^s(\boldsymbol{\Lambda}_{sol}))$ 
and \beq \label{eq:Phi_I_S}
\boldsymbol{\Phi}_{I}:=[\boldsymbol{\lambda}_I, \boldsymbol{\lambda}_I^2,\ldots, \boldsymbol{\lambda}_I^{K_l}], \;  \boldsymbol{\Phi}_{s}:=[\boldsymbol{\lambda}_s, \boldsymbol{\lambda}_s^2,\ldots, \boldsymbol{\lambda}_s^{K_u}], \eeq with $\boldsymbol{\lambda}_I$ and $\boldsymbol{\lambda}_s$ the column vectors  with entries, respectively,  the non-zero eigenvalues of the lower and upper Laplacians.
The optimal solutions of the two least-squares problems in 
 (\ref{eq:H_irr1})  are given in closed form by
 \beq
 \ba^{I}=\boldsymbol{\Phi}_{I}^{\dag} \mathbf{h}^{I}, \qquad \ba^{s}=\boldsymbol{\Phi}_{s}^{\dag} \mathbf{h}^{s}
 \eeq
where  $\boldsymbol{\Phi}^{\dag}$ denotes the Moore-Penrose pseudo-inverse of $\boldsymbol{\Phi}$.
Note that, defining 
  $\mathbf{h}=[\mathbf{h}^{s};\mathbf{h}^{I}]$ and $\boldsymbol{\Phi}_{t}=[\boldsymbol{\Phi}_{s};\boldsymbol{\Phi}_{I}]$, a single filter can be derived by solving the following least-squares problem
\beq \label{eq:H_g1}
\underset{\ba \in \mathbb{R}^{K}}{\min} \quad \parallel \bh-\boldsymbol{\Phi}_t {\ba} \parallel^2_F \eeq
    whose optimal solution is given in closed form by
$\ba=\boldsymbol{\Phi}^{\dag}_t \bh.$

\begin{figure}[ht]
\centerline{\includegraphics[width=8.4cm,height=5cm]{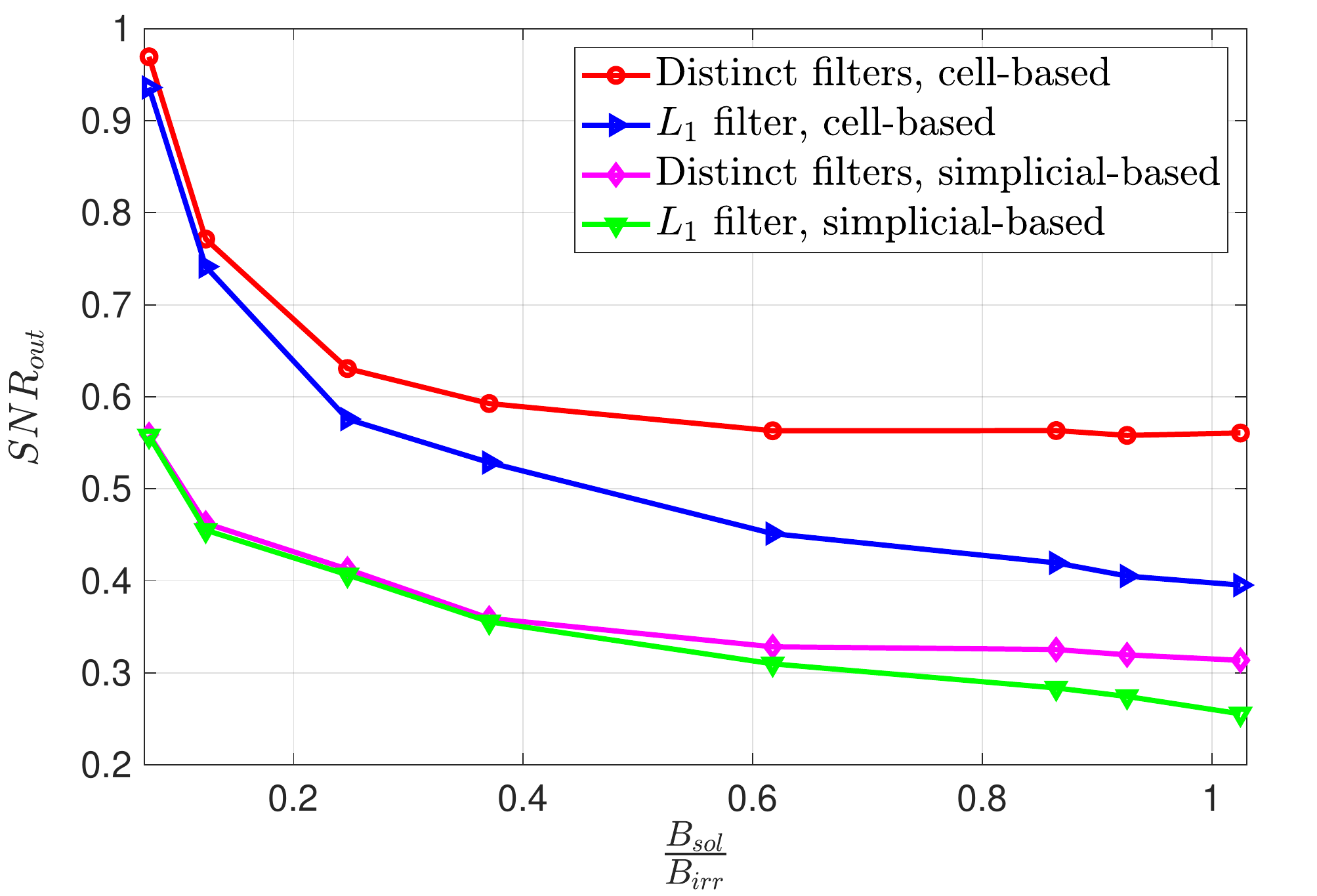}}
\caption{$\text{SNR}$ at the filter output for the solenoidal part vs.   $B_{sol}/B_{irr}$.}
\label{fig: fig_SNR}
\end{figure}

%As we will numerically assess, the independent design of the two filters coefficients leads  performance gains in term of filter approximation error with respect to a joint filter design.
In Fig. $3$ to evaluate the goodness of the proposed filtering strategy we report the signal-to-noise ratio (SNR) observed at the output of the solenoidal filter versus the ratio between the solenoidal and irrotational signal bandwidths.
We consider the optimal FIR filters for the solenoidal and  irrotational filters obtained by solving the problems in (\ref{eq:H_irr1})  or by using a single $\mL_1$-based FIR filter as solution of the problem in (\ref{eq:H_g1}). It can be observed from Fig. $3$ that the independent design of the two filters coefficients provides a performance gains  with respect to a single $\mL_1$-based filter and to simplicial-based methods.
%\vspace{-0.1cm}

\section{Conclusions}
In this paper we extended topological signal processing from simplicial complexes to richer topological spaces as cell complexes.  We show as TSP over cell complex  enables sparser signal representations than simplicial complexes based approaches. Furthermore, we focused on the optimal design of FIR filters for the solenoidal and irrotational signals to minimize the approximation error with respect to the desired spectral masks. 
%\section*{Acknowledgment}
%\vspace{-0.1cm}

%\section*{References}

\bibliographystyle{IEEEbib}
\bibliography{reference}
\end{document}